\documentclass[epj]{svjour}
\usepackage{graphicx,wrapfig,lipsum}
\usepackage{amssymb,amsmath}
\usepackage{subeqnarray}
\usepackage{url,hyperref,doi} 
\makeatletter
\g@addto@macro{\UrlBreaks}{\UrlOrds}
\makeatother

\newcommand{\req}[1]{Eq.\,(\ref{#1})} 
\newcommand{\Req}[1]{Equation\,(\ref{#1})}

\newcommand{\ie}{{\it i.e.\/}}

\newcommand{\eg}{{\it e.g.\/}}

\newcommand{\EE}{\ensuremath{\,{\vec {\cal E}}}}
\newcommand{\BB}{\ensuremath{\,{\vec {\cal B}}}}
\newcommand{\TT}{\ensuremath{\,{\vec {\cal T}}}}
\newcommand{\FF}{\ensuremath{\,{\vec {\cal F}}}}

\newcommand{\lambdabar}{\ensuremath{\mkern0.75mu\mathchar '26\mkern -9.75mu\lambda}}

\newcommand{\dsfrac}{\displaystyle\frac}

\begin{document}
\title{Relativistic Dynamics of Point Magnetic Moment}
\author{Johann Rafelski, Martin Formanek, and Andrew Steinmetz}
%
%
\institute{Department of Physics, The University of Arizona, 
Tucson, Arizona, 85721, USA}
\date{Submitted:December 1, 2017 / Print date: \today}
%
\abstract{The covariant motion of a classical point particle with magnetic moment in the presence of (external) electromagnetic fields is revisited. We are interested in understanding Lorentz force extension involving point particle magnetic moment (Stern-Gerlach force) and how the spin precession dynamics is modified for consistency. We introduce spin as a classical particle property inherent to Poincare\'e symmetry of space-time. We propose a covariant formulation of the magnetic force based on a \lq magnetic\rq\ 4-potential and show how the point particle magnetic moment relates to the Amperian (current loop) and Gilbertian (magnetic monopole) description. We show that covariant spin precession lacks a unique form and discuss connection to $g-2$ anomaly. We consider variational action principle and find that a consistent extension of Lorentz force to include magnetic spin force is not straightforward. We look at non-covariant particle dynamics, and present a short introduction to dynamics of (neutral) particles hit by a laser pulse of arbitrary shape.
} 
\maketitle

\section{Introduction}
The (relativistic) dynamics of particle magnetic moment $\vec{\mu}$, \ie\ the proper time dynamics of spin $ s^\mu(\tau)$, has not been fully described before. Our interest in this topic originates in a multitude of current research topics:\\ 
i) the ongoing effort to understand the magnetic moment anomaly of the muon~\cite{Grange:2015fou,Bennett:2006fi};\\
ii) questions regarding how elementary magnetic dipoles (\eg\ neutrons) interact with external fields~\cite{2014McDonaldNeutron,1986Mezei};\\
iii) particle dynamics in ultra strong magnetic fields created in relativistic heavy ion collisions~\cite{Huang:2015oca,Greif:2017irh};\\
iv) magnetars, stellar objects with extreme ${\cal O}(10^{11})\,\mathrm{T}$ magnetic fields~\cite{Turolla:2015mwa,Sedrakian:2017lpy};\\
v) the exploration of particle dynamics in laser generated strong fields~\cite{Wen:2017zer};\\
vi) neutron beam guidance and neutron storage rings~\cite{Kugler:1978ae}; 
and\\
vii) the finding of unusual quantum spin dynamics when gyromagnetic ratio $g\ne 2$~\cite{Rafelski:2012ui,AndrewSCoulomb}.\\
The results we present will further improve the understanding of plasma physics in presence of inhomogeneous magnetic fields, and improve formulation of radiation reaction forces, topics not further discussed in this presentation.

In the context of the electromagnetic (EM) Maxwell-Lorentz theory we learn in the classroom that 
\begin{enumerate}
\item 
The magnetic moment $\vec{\mu}$ has an interaction energy with a magnetic field $ \BB$ 
\begin{equation}\label{muB}
E_m=-\vec \mu \cdot \BB\;.
\end{equation}
The corresponding Stern-Gerlach force $ \FF_\mathrm{SG}$ has been written in two formats
\begin{equation}\label{SGF}
\FF_\mathrm{SG}\equiv
\left\{\begin{array}{lr}
 \vec \nabla (\vec \mu\cdot \BB ) \;, &\ \mathrm{Amperian\ Model}\;, \\[0.2cm]
 (\vec \mu\cdot \vec \nabla ) \BB \;, & \mathrm{Gilbertian\ Model}\;.
\end{array}
\right.
\end{equation} 
The name \lq Amperian\rq\ relates to the loop current generating the force. The Gilbertian model invokes a magnetic dipole made of two magnetic monopoles. These two forces written here in rest frame of a particle are related~\cite{2014McDonaldNeutron,1986Mezei}. We will show that a internal spin based magnetic dipole appears naturally; it does not need to be made of magnetic monopoles or current loops. We find that both force expressions in \req{SGF} are equivalent, this equivalence arises from covariant dynamics we develop and requires additional terms in particle rest frame complementing those shown in \req{SGF}.
\item
The torque \TT\ that a magnetic field $\BB$ exercises on a magnetic dipole $\vec{\mu}$ in a way that tends to align the dipole with the direction of a magnetic field $\BB$
\begin{equation}\label{tork}
\TT \equiv \dsfrac{d\vec{s}}{dt}=\vec \mu \times \BB 
= g \mu_\mathrm{B} \;\frac{\vec s}{\hbar/2}\times \BB\;, \quad \mu_\mathrm{B}\equiv \dsfrac{e\hbar}{2m}\;.
\end{equation}
The magnetic moment is defined in general in terms of the product of Bohr magneton $\mu_\mathrm{B}$ with the gyromagnetic ratio $g$, $|\mu|\equiv g\mu_\mathrm{B}$. In \req{tork} we used $|\vec s|=\hbar/2$ for a spin-1/2 particle, a more general expression will be introduced in subsection~\ref{ssec:dipolepot}.
\end{enumerate} 
 
We used the same coefficient $\vec \mu$ to characterize both the Stern-Gerlach force \req{SGF} and spin precession force \req{tork}. However, there is no compelling argument to do so and we will generalize this hypothesis --- it is well known that Dirac quantum dynamics of spin-1/2 particles predicts both the magnitude $g=2$ and identity of magnetic moments entering \req{SGF} and \req{tork}. 

While the conservation of electrical charge is rooted in gauge invariance symmetry, the magnitude of electrical charge has remained a riddle; the situation is similar for the the case of the magnetic moment $\vec \mu$: spin properties are rooted in the Poincar\'e symmetry of space-time, however, the strength of spin interaction with magnetic field, \req{muB} and \req{tork}, is arbitrary but unique for each type of (classical) particle. Introducing the gyromagnetic ratio $g$ we in fact create an additional conserved particle quality. This becomes clearer when we realize that the appearance of \lq $e$\rq\ does not mean that particles we study need to be electrically charged.

First principle considerations of point particle relativistic dynamics experience some difficulties in generating Eqs.\,(\ref{SGF},\ref{tork}), as a rich literature on the subject shows -- we will cite only work that is directly relevant to our approach; for further 70+ references see the recent numerical study of spin effects and radiation reaction in a strong electromagnetic field~\cite{Wen:2017zer}.

For what follows it is important to know that the spin precession \req{tork} is a result of spatial rotational invariance which leads to angular and spin coupling, and thus spin dynamics can be found without a new dynamical principle has been argued \eg\ by Van Dam and Ruijgrok~\cite{VanDam:1980aw} and Schwinger~\cite{Schwinger:1974as}. Similar physics content is seen in the work of Skagerstam and Stern~\cite{Skagerstam:1981xp,Balachandran:2017jha}, who considered the context of fiber bundle structure focusing on Thomas precession.
 
Covariant generalization of the spin precession \req{tork} is often attributed to the 1959 work by Bergmann-Michel-Telegdi~\cite{Bargmann:1959gz}. However we are reminded~\cite{Jackson2008,hushwaterJDJComm,hushwaterJDJCommR} that this result was discovered already 33 years earlier by L.H. Thomas~\cite{Thomas:1926dy,Thomas:1927yu} at the time when the story of the electron gyromagnetic ratio $g=2$ was unfolding. Following Jackson~\cite{Jackson2008} we call the corresponding equation TBMT. J. Frenkel, who published~\cite{1Frenkel:1926zz,Frenkel:1926zz} at the same time with L.H. Thomas explored covariant form of the Stern-Gerlach force, a task we complete in this work. 

There have been numerous attempts to improve the understanding of how spin motion back-reacts into the Lorentz force generating the the Stern-Gerlach force. In the 1962 review Nyborg~\cite{P. Nyborg} summarized efforts to formulate the covariant theory of electromagnetic forces including particle intrinsic magnetic moment. In 1972 Itzykson and Voros~\cite{Itzykson:1972py} proposed a covariant variational action principle formulation introducing the inertia of spin $I$, seeking consistent variational principle but they found that no new dynamical insight resulted in this formulation.

Our study relates most to the work of Van Dam and Ruijgrok~\cite{VanDam:1980aw}. This work relies on an action principle and hence there are in the Lorentz force inconsistent terms that violate the constraint that the speed of light is constant, see \eg\ their Eq.\,3.11 and remarks: \lq\ The last two terms are ${\cal O}(e^2)$ and will be omitted in what follows.\rq\ Other authors were proposing mass modifications to compensate these terms, a step which is equally unacceptable. For this reason our approach is intuitive, without insisting on \lq in principle there is an action\rq. Once we have secured a consistent, unique covariant extension of the Lorentz force, we explore the natural variational principle action. We find it is not consistent and we identify the origin of the variational principle difficulties.

We develop the concept of the classical point particle spin vector in the following section~\ref{SpinV}. Our discussion relates to Casimir invariants rooted in space-time symmetry transformations. Using Poincar\'e group generators and Casimir eigenvalues we construct the particle momentum $p^\mu$ and particle space-like spin pseudo vector $s^\mu$. In section~\ref{Pdynamics} we present a consistent picture of the Stern-Gerlach force (subsection~\ref{LMotion}) and generalize TBMT precession equation (subsection~\ref{SMotion}) linear in both, the EM field and EM field derivatives. We connect the Amperian form of SG force (\ref{ssec:dipolepot}) with the Gilbertian force (\ref{ssec:GilbertF}). We discuss non-uniqueness of spin dynamics (\ref{ssec:TBMTgradNU}) in consideration of impact on muon $g-2$ experiment. We show in section~\ref{VarPrin} that the natural choice of action for the considered dynamical system does not lead to a consistent set of equations; in this finding we align with all prior studies of Stern-Gerlach extension of the Lorentz force. 

In the final part of this work section~\ref{Experiment} we show some of the physical content of the theoretical framework. In subsection~\ref{NonCov} we present a more detailed discussion of dynamical equations for the case of particle in motion with a given $\vec \beta=\vec v/c$ and \EE,\ \BB\ fields in laboratory. In subsection~\ref{PlaneWM} we study solution of the dynamical equations for the case of an EM light wave pulse hitting a neutral particle. We have obtained exact solutions of this problem, detail will follow under seperate cover~\cite{FormanekSOl}. The concluding section~\ref{conclude} is a brief summary of our findings.

\subsection{Notation}
For most of notation, see Ref.\,\cite{Rafelski:2017hyt}. Here we note that we use SI unit system and the metric:
$$
\mathrm{diag}\; g_{\mu\nu}=\{1,-1,-1,-1\}\;, \
p_\mu p^\mu= g_{\mu\nu}p^\mu p^\nu=\frac{E^2}{c^2}-\vec p^{\,2}\;,
$$
We further recognize the totally antisymmetric covariant pseudo tensor $\epsilon$:
$$
\epsilon_{\mu\nu\alpha\beta}=\sqrt{-g}
\left\{
\begin{array}{lr}
(-1)^\mathrm{perm}\;, & \quad\mathrm{if\ all\ indices\ are\ distinct}\\[0.1cm]
0\;, & \mathrm{otherwise}\;,
\end{array}
\right.
$$
where \lq perm\rq\ is the signature of the permutation. It is important to remember when transiting to non-covariant notation in laboratory frame of reference that the analog contravariant pseudo-tensor due to odd number of space-like dimensions is negative for even permutations and positive for odd permutations. The Appendix B of Ref.\,\cite{Wald:1984rg} presents an introduction to $\epsilon$.

We will introduce an elementary magnetic dipole charge $d$ --- the limitations of the alphabet force us to adopt the letter $d$ otherwise used to describe the electric dipole to be the elementary magnetic dipole charge. The magnetic dipole charge of a particle we call $d$ converts the spin vector $\vec s$ to magnetic dipole vector $\vec \mu$,
\begin{equation}\label{defd}
\vec s dc= \vec \mu \;,\qquad d\equiv \dsfrac{|\vec{\mu}|}{c|\vec{s}|}\;.
\end{equation}
The factor $c$ is needed in SI units since in the EM-tensor $F^{\mu\nu}$ has as elements $\EE/c$ and $\BB$. It seems natural to introduce also $s^\mu d=\mu^\mu$ --- an object $\mu^\mu$ can confuse and we stick to the product $s^\mu d$, however we always replace $\vec s d\to \vec \mu/c$. Note that we place $d$ to right of pertinent quantities to avoid confusion such as $dx$.

We cannot avoid appearance in the same equation of both magnetic moment $\vec \mu$ and vacuum permeability $\mu_0$. 

\section{Spin Vector} \label{SpinV}
A classical intrinsic covariant spin has not been clearly defined or even identified in prior work. In some work addressing covariant dynamics of particles with intrinsic spin and magnetic moment particle spin is by implication solely a quantum phenomenon. Therefore we describe the precise origin of classical spin conceptually and introduce it in explicit terms in the following. 

Considering the Poincare group of space-time symmetry transformations~\cite{Greiner:1984uv,Ohlsson:2011zz}, it has been established that elementary particles have to be in a representation that is characterized by eigenvalues of two Casimir operators
(a \lq bar\rq\ marks operators)
\begin{equation}\label{Casimir}
\bar C_1\equiv \bar p_\mu \bar p^\mu=\bar p^2\equiv m^2c^2\;,\quad 
\bar C_2\equiv \bar w_\alpha \bar w^\alpha\;.
\end{equation}
All physical point \lq particles\rq\ have fixed eigenvalues of $C_1, C_2$.
The quantities (with a bar) $\bar p^\mu$ and $\bar w^\alpha$ are differential operators constructed from generators of the symmetry transformations of space-time; that is 10 generators of the Poincar\'e group of symmetry transformations of 4-space-time: $\bar p^\mu$ for translations, $\overline{\vec J}$ for rotations and $\overline{\vec K}$ for boosts. Once we construct suitable operator valued quantities we will transition to the physics of \lq c-number\rq\ valued (without bar) variables as used in classical dynamics where all quantities will be normal numbers and rely on the eigenvalues of Casimir operators $C_1,C_2$ for each type of particle.

In \req{Casimir} the first of the space-time operators based on generators of the four space-time translations $p^\mu$ guarantees that a point particle has a conserved inertial mass $m$ (with a value specific for any particle type). The second Casimir operator $C_2$ is obtained from the square of the Pauli-Luba\'nski 4-vector 
\begin{equation}\label{PLvector}
\bar w_\alpha=\overline M^\star_{\alpha\beta}\bar p^\beta\;
\qquad
\overline M^\star_{\alpha\beta}\equiv \frac 1 2 \epsilon_{\alpha\beta\mu\nu}\overline M^{\mu\nu}\;. 
\end{equation}
Here $\overline M^{\mu\nu}$ is the antisymmetric tensor (operator) created from three Lorentz-boost generators $\overline{\vec K}$ and three space rotation generators $\overline{\vec J}$ such that 
\begin{equation}\label{MKJ}
\frac 1 2 \overline M_{\mu\nu}\overline M^{\nu\mu}=\overline{\vec K}^{\,2}-\overline{\vec J}^{\,2}\;,
\qquad
\frac 1 4 \overline M^\star_{\mu\nu}M^{\mu\nu} =\overline{\vec J}\cdot\overline{\vec K}\;.
\end{equation}
These relations help us see that 
$$F^{\mu\nu}(\vec E\to\overline{\vec K}\,,\vec B\to\overline{\vec J})=\overline M^{\mu\nu}\;.$$
The generators $\overline{\vec J},\;\overline{\vec K}$ of space-time transformations are recognized by their commutation relations. They are used in a well known way to construct representations of the Lorentz group.

In terms of the generator tensor $\overline M^{\nu\mu}$ the covariant definition of the particle spin (operator) vector is 
\begin{equation}\label{sdef}
\bar s_\mu\equiv \dsfrac{\bar w_\mu}{\sqrt{C_1}}= \overline M^\star_{\mu\nu}\,\frac{\bar u^\nu}{c}\;, \quad 
\bar u^\mu\equiv \dsfrac{c\bar p^\mu}{ \sqrt{C_1}}= \dsfrac{\bar p^\mu}{m}\;.
\end{equation}
According to \req{sdef}, spin $\bar s_\mu$ is a pseudo vector, as required for angular dynamics. The dimension of $\bar s^\mu$ is the same as the dimension of the generator of space rotations $\overline{\vec J}$. We further find that $\bar s^\mu$ is orthogonal to the 4-velocity (operator) $\bar u^\mu$
\begin{equation}\label{sdef2}
c\bar s \cdot\bar u =\bar u^\nu \overline M^\star_{\nu\mu}\bar u^\mu=0\;,
\end{equation}
by virtue of the antisymmetry of $\overline M^\star$ evident in the definition \req{PLvector}. The definition of the particle spin (operator) is unique: no other space-like (space-like given the orthogonality $\bar s\cdot \bar u=0$) pseudo vector associated with the Poincar\'e group describing space-time symmetry transformations can be constructed. 

We now transition to c-numbered quantities (dropping the bar): an observer \lq (0)\rq\ co-moving with a particle measures the 4-momentum and 4-spin $s^\mu$
\begin{equation}\label{RFrame} 
p_{(0)}^\mu\equiv \{\sqrt{C_1},0,0,0\}\;,\quad
s_ {(0)}^\mu\equiv \{0,0,0,\sqrt{|C_2|/C_1}\}
\end{equation}
where according to convention $\hat z$-axis of the coordinate system points in direction of the intrinsic spin vector $\vec s$. In the particle rest frame we see that 
\begin{equation}\label{constraint} 
0=p_{(0)}^\mu s^{(0)}_ {\, \mu} =p^\mu s_ {\mu}|_{(0)} = m(u^\mu s_\mu) |_\mathrm{any\ frame}\;,
\end{equation}
consistent with operator equation \req{sdef2}; more generally, any space-like vector is normal to the time-like 4-velocity vector. For the magnitude of the spin vector we obtain
\begin{equation}\label{RFrame1} 
-\vec s^{\,2}\equiv s^\mu_{(0)} s^{(0)}_{\,\mu} \equiv s^\mu s_\mu|_{(0)}=s^2|_\mathrm{any\ frame}= \dsfrac{-|C_2|}{C_1}\;.
\end{equation}
We keep in mind that $s^2$ must always be a constant of motion in any frame of reference. Its value $s\cdot s=-\vec s^{\,2}$ is always negative, appropriate for a space-like vector. Similarly
\begin{equation}\label{RFrame1P} 
p_{(0)}^\mu p^{(0)}_\mu =p^2|_\mathrm{any\ frame} = C_1 \equiv m^2c^2 \;,
\end{equation}
must be a constant of motion in any frame of reference and the value $p^2$ is positive, appropriate for a time-like vector. 

As long as forces are small in the sense discussed in Ref.\,\cite{Rafelski:2017hyt} we can act as if rules of relativity apply to both inertial and (weakly) accelerated frames of reference. This allows us to explore the action of forces on particles in their rest frame where \req{RFrame} defines the state of a particle. By writing the force laws in covariant fashion we can solve for dynamical evolution of $p^\mu(\tau),\;s^\mu(\tau)$ classical numbered variables.

\section{Covariant dynamics}\label{Pdynamics}
\subsection{Generalized Lorentz force}\label{LMotion}
\subsubsection{Magnetic dipole potential and Amperian force} \label{ssec:dipolepot}
We have gone to great lengths in section~\ref{SpinV} to argue for the existence of particle intrinsic spin. For all massive particles this implies the existence of a particle intrinsic magnetic dipole moment, without need for magnetic monopoles to exist or current loops. Spin naturally arises in the context of symmetries of Minkowski space-time, it is not a quantum property. 

In view of above it is appropriate to study classical dynamics of particles that have both, an elementary electric charge $e$, and an elementary magnetic dipole charge $d$. The covariant dynamics beyond the Lorentz force needs to incorporate the Stern-Gerlach force. Thus the extension has to contain the elementary magnetic moment of a particle contributing to this force. To achieve a suitable generalization we introduce the magnetic potential 
 \begin{equation}\label{genB2}
\boxed{
B_\mu(x\,,s)\,d \equiv F^\star_{\mu\nu}(x)s^\nu\,d\;,
}\quad 
F^\star_{\mu\nu}=\frac 1 2 \epsilon_{\mu\nu\alpha\beta}F^{\alpha\beta}\;.
\end{equation}
We use dual pseudo tensor 
since $s_\mu$ is a pseudo vector; the product in \req{genB2} results in a polar 4-vector $B_\mu$. We note that the magnetic dipole potential $B_\mu$ by construction in terms of antisymmetric field pseudo tensor $F^\star_{\mu\nu}$ satisfies
\begin{equation}\label{genB3}
\partial_\mu B^\mu=0\;,\qquad
s\cdot B=0\;, \to \quad B\cdot \frac{ds}{d\tau}=-s\cdot \frac{dB}{d\tau}\;.
\end{equation}

The additional potential energy of a particle at rest placed in this magnetic dipole potential is 
\begin{equation}\label{PotMu}
U_{(0)}\equiv B^0\,c\,d =cF^\star_{0\nu}(x)s^\nu d=-|\vec{\mu}|\,\BB\cdot \frac{\vec s}{|\vec{s}|}\equiv -\vec\mu\cdot \BB\;.
\end{equation}
This shows \req{genB2} describes the energy content seen in \req{muB}; all factors are appropriate.

The explicit format of this new force is obtained when we use \req{genB2} to define a new antisymmetric tensor
\begin{equation}\label{SGforce}
G^{\mu\nu} =\partial^\mu B^\nu-\partial^\nu B^\mu
= s_\alpha\left[\partial^\mu F^{\star\,{\nu\alpha}}-
 \partial^\nu F^{\star\,{\mu\alpha}}\right]\;.
\end{equation}
\Req{SGforce} allows us to add to the Lorentz force 
\begin{equation}\label{SGforce2}
m\dot u^\mu= H^{\mu\nu}u_\nu\;,\quad
H^{\mu\nu}= eF^{\mu\nu} + G^{\mu\nu}\,d\;.
\end{equation}
In the $G$-tensor we note appearance in the force of the derivative of EM fields, required if we are to see the Amperian model variant of the Stern-Gerlach force \req{SGF} as a part of generalized Lorentz force. 

The Amperian-Stern-Gerlach (ASG) force 4-vector is obtained multiplying with $u_\nu d$ the $G$-tensor \req{SGforce}. Thus the total 4-force a particle of charge $e$ and magnetic dipole charge $d$ experiences is 
\begin{equation}\label{Aforce1}
\boxed{
F^\mu_\mathrm{ASG}=
eF^{\mu\nu}u_\nu -u\cdot \partial\, F^{\star\,\mu\nu}s_\nu\, d +
\partial^\mu (u\cdot\, F^\star\cdot s \,d)\;.
}
\end{equation}
In the particle rest frame we have 
\begin{equation}\label{usRF}
u^\nu|_\mathrm{RF}=\{c,\vec 0\}\;,\quad cs^\nu d|_\mathrm{RF}=\{0,\vec\mu\}\;.
\end{equation}
We can use \req{usRF} to read-off from \req{SGforce2} the particle rest frame force to be
\begin{equation}\label{Aforce2}
F^\mu_\mathrm{ASG}|_\mathrm{RF}=\left\{0,\;e\EE
-\dsfrac{1}{c^2}\,\vec \mu \times \dsfrac{\partial \EE}{\partial t}
+\vec \nabla(\vec \mu\cdot \BB)\right\}\;,
\end{equation}
where two contributions $\partial(\vec \mu\cdot \BB)/\partial t$ to $F^0$ cancel. Each of the three terms originates in one of the covariant terms in the sequence shown. The result is what one calls Amperian model originating in dipoles created by current loops. This is, however, not the last word in regard to the form of the force. 

\subsubsection{Gilbertian model Stern-Gerlach force} \label{ssec:GilbertF}
We restate the Stern-Gerlach-Lorentz force \req{SGforce2}, showing the derivative terms explicitly, 
\begin{equation}\label{SGforce20}
m\dot u^\mu=e F^{\mu\nu}u_\nu+
\left(\partial^\mu (u\cdot F^\star\!\cdot s)
-s_\alpha u\cdot \partial F^{\star\,{\mu\alpha}}\right)\,d\;.
\end{equation}
Multiplying with $s^\mu$ the last term vanishes due to antisymmetry of $F^\star$ and we obtain 
\begin{equation}\label{SGforce21}
s\cdot \dot u=\frac{1}{m}s\cdot \left(eF-
 s\cdot \partial\, F^\star\,d\right)\cdot u
\;.
\end{equation}
This equation suggests that we explore
\begin{equation}\label{FtoFtilde} 
eF^{\mu\nu}\to \ \boxed{\widetilde F^{\mu\nu}= eF^{\mu\nu}\! 
 - s\cdot \partial\, F^{\star\,{\mu\nu}}\,d}\;,
\end{equation}
as the generalized Lorentz force replacing the usual field tensor $eF$ by $\widetilde F$ in a somewhat simpler way compared to the original $H^{\mu\nu}$ \req{SGforce2} modification. 

We demonstrate now that the field modification seen in \req{FtoFtilde} leads to a different and fully equivalent format of the force. We replace in the first term in \req{SGforce20} $F\to \widetilde F$ and add the extra term from \req{FtoFtilde} to the two reminder terms. Changing the index naming these we can write symmetrically
\begin{align}\label{SGforceF}
m\dot u^\mu=& \widetilde F^{\mu\nu}u_\nu\\ \notag
+&s_\alpha \left(
\partial^\alpha F^{\star\,{\mu\beta}} +
\partial^\mu F^{\star\,{\beta\alpha}} +
\partial^\beta F^{\star\,{\alpha\mu}} 
\right) u_\beta \, d\;.
\end{align}
The tensor appearing in the parentheses in the 2nd line of \req{SGforceF} is antisymmetric under any of the three exchanges of the indices. It is therefore proportional to the totally antisymmetric tensor $\epsilon^{\alpha\mu\beta\gamma}$ which must be contracted with some 4-vector $V_\gamma$ containing a gradient of the EM dual field tensor, there are two such available 4-vectors 
$\partial^\kappa F^\star_{\kappa\gamma}$ which vanishes by virtue of Maxwell equations, and 
$$
V_\gamma = \frac 1 2 \epsilon_{\gamma\kappa\eta\zeta}\partial^\kappa F^{\star\,{\eta\zeta}}=\partial^\kappa F_{\kappa\gamma}=\mu_0 j_\gamma\;.$$ 
Thus we introduce the Gilbertian form of the 4-force
\begin{equation}\label{SGforce200}
\boxed{
F_\mathrm{GSG}^\mu=\widetilde F^{\mu\nu}u_\nu-\mu_0 j^\gamma \epsilon_{\gamma\alpha\beta\nu} u^\alpha s^\beta g^{\nu\mu}\, d 
\;.}
\end{equation}
Note that in our formulation the Amperian and the Gilbertian 4-forces are identical
\begin{equation}\label{SGforce230}
F_\mathrm{ASG}^\mu=F_\mathrm{GSG}^\mu\;,
\end{equation}
they are just written differently.

In the rest frame of a particle, see \req{usRF} the Gilbertian force \req{SGforce230} is 
\begin{equation}\label{SGforceFinalRest}
F_\mathrm{GSG}^\mu|_\mathrm{RF}=\left\{0,\;e\EE+(\vec \mu\cdot \vec \nabla ) \BB+\mu_0 \vec \mu \times \vec j\right\}\;.
\end{equation}
It is interesting to see the mechanism by which the two formats of the forces equal to each other in the particle rest frame. With
$$
 \vec \nabla(\vec \mu\cdot \BB )-(\vec \mu\cdot \vec \nabla ) \BB=\vec \mu\times (\vec \nabla\times \BB)\;,
$$
we form the force difference between \req{Aforce2} and \req{SGforceFinalRest}
\begin{equation}\label{SGforceFinalRest1}
\left[\vec F_\mathrm{ASG}-\vec F_\mathrm{GSG}\right]_\mathrm{RF}\!\!\! = \vec \mu\times \left(\!
-\frac{1}{c^2}\dsfrac{\partial\EE}{\partial t}
\!+\vec \nabla\times \BB 
- \mu_0\vec j\right)\!=0
\;.
\end{equation}
The terms in parenthesis cancel according to Maxwell equation confirming that both the Amperian and the Gil\-bertian forces are equal taking as an example the instantaneous rest frame. From now one we will use Gilbertian form of the force and in later examples we will focus on particle motion in vacuum, $j^\mu=0$.

In this discussion of forces we kept the electrical charge $e$ and the elementary magnetic moment \lq charge\rq\ $d$ \req{defd} as independent qualities of a point particle. As noted in the introduction it is common to set $ |\vec \mu|\equiv g\mu_\mathrm{B}$, see below \req{tork}. Hence we can have both, charged particles without magnetic moment, or neutral particles with magnetic moment, aside from particles that have both charge and magnetic moment. For particles with both charge and magnetic moment we can write, using Gilbertian format of force
\begin{equation}\label{SGforceFinal}
m\dot u^\mu= \widetilde F^{\mu\nu}u_\nu =e\left(F^{\mu\nu}-(1+a)\,\lambdabar\frac{s\cdot \partial}{|\vec{s}|} \;F^{\star\,{\mu\nu}}\right)u_\nu\;,
\end{equation}
where $a=(g-2)/2$ is the gyromagnetic ratio anomaly. The Compton wavelength $\lambdabar = {\hbar}/{mc}$ defines the scale at which the spatial field inhomogeneity is relevant; note that inhomogeneities of the field are boosted in size for a particle in motion, a situation which will become more explicit in section~\ref{ssec:EfieldTOP}.

\subsection{Spin motion}\label{SMotion}
\subsubsection{Conventional TBMT}\label{sssec:TBMT}
For particles with $m\ne 0$ differentiating \req{constraint} with respect to proper time we find 
\begin{equation}\label{spinM}
\dot u\cdot s+u\cdot \dot s=0\;, 
\end{equation}
where we introduced proper time derivative $\dot s^\mu=d s^\mu/d\tau$. Schwinger observed~\cite{Schwinger:1974as} that given \req{spinM} one can use covariant form of the dynamical Lorentz force equations for ${du^\mu}/{d\tau}$ to obtain
\begin{equation}\label{Schwinger1}
 u_\mu\left(\frac{ds^\mu}{d\tau}-\frac e m F^{\mu\nu}s_\nu\right)=0\;.
\end{equation}
Here $F^{\mu\nu}$ is the usual EM field tensor. \Req{Schwinger1} has the general TBMT solution
\begin{equation}\label{TBMT} 
\frac{ds^\mu}{d\tau}=\frac e m F^{\mu\nu}s_\nu
+ \frac{\widetilde ae}{m} \left( F^{\mu\nu}s_\nu- \dsfrac{u^\mu}{c^2} (u\cdot F \cdot s)\right)\;,
\end{equation}
where we used the notation $u\cdot F\cdot s\equiv u_\mu F^{\mu\nu}s_\nu\;$. 

In \req{TBMT} $\widetilde a$ is an arbitrary constant considering that the additional term multiplied with $u^\mu$ vanishes. On the other hand we can read off the magnetic moment entering \req{tork}: the last term is higher order in $1/c^2$. Hence in the rest frame of the particle we see that $2(1+\widetilde a)=g$ \ie\ \req{TBMT} reproduces \req{tork} with the magnetic moment coefficient when $\widetilde a = a$. Therefore, as introduced, $\widetilde a=a$ is the $g\ne 2$ anomaly. However, in \req{TBMT} we could for example use $\widetilde a = (g^2-4)/8= a+a^2/2$, which classical limit of quantum dynamics in certain specific conditions implies~\cite{AndrewSCoulomb}. In this case $\widetilde a\to a$ up to higher order corrections. This means that measurement of $\widetilde a$ as performed in experiments~\cite{Grange:2015fou,Bennett:2006fi} depends on derivation of the relation of $\widetilde a$ with $a$ obtained from quantum theory. These remarks apply even before we study gradient in field corrections.

\subsubsection{Gradient corrections to TBMT}\label{ssec:TBMTgrad}
The arguments by Schwinger, see Eqs.\,(\ref{spinM},\ref{Schwinger1},\ref{TBMT}), are ideally positioned to obtain in a consistent way generalization of the TBMT equations including the gradient of fields terms required for consistency. We use \req{FtoFtilde} in \req{TBMT} to obtain
\begin{align}\label{TBMTgradFinal} 
\frac{ds^\mu}{d\tau}=&\; \frac{ 1+\widetilde a }{m}\left(\;e\,F^{\mu\nu}\;-
 \; s\cdot \partial\, \;\; F^{\star\,{\mu\nu}}\;d \,\right)s_\nu\\ \notag
+& \ \ \frac{\widetilde a}{mc^2}\left(\!s\!\cdot\! eF\!\cdot u 
 - s\cdot \partial \,s\!\cdot F^*\!\!\!\cdot u\,d\right)u^\mu \;.
\end{align} 

The dominant gradient of field correction arises for an elementary particle from the 2nd term in the first line in \req{TBMTgradFinal}, considering the coefficient of the second line $a =\alpha_2/2\pi+\ldots=1.2\times 10^{-3}$. One should remember that given the precision of the measurement~\cite{Grange:2015fou,Bennett:2006fi} of $\widetilde a$ which is driven by the first term in the second line in \req{TBMTgradFinal} we cannot in general neglect the new 2nd term in first line in \req{TBMTgradFinal} even if the characteristic length defining the gradient magnitude is the Compton wavelength $\lambdabar$, see \req{SGforceFinal}.

\subsubsection{Non-uniqueness of gradient corrections to TBMT}\label{ssec:TBMTgradNU}
It is not self evident that the form \req{TBMTgradFinal} is unique. To see that a family of possible extensions TBMT arises we recall the tensor \req{SGforce2} $H^{\mu\nu}$ made of the two potentials $A^\mu$ and $B^\mu$. We now consider the spin dynamics in terms of the two field tensors, $F$ and $G$ replacing the usual EM-tensor $F^{\mu\nu}$in the Schwinger solution, \req{TBMT}. In other words, we explore the dynamics according to 
\begin{align}\label{TBMT2}
\frac{ds^\mu}{d\tau}=&\frac 1 m eF^{\mu\nu}s_\nu
+ \frac {\widetilde a e}{ m} \left(F^{\mu\nu}s_\nu- \dsfrac{u^\mu}{c^2} (u\cdot F \cdot s)\right)\\ \notag
+&G^{\mu\nu}s_\nu\frac d m 
+ \left(G^{\mu\nu}s_\nu- \dsfrac{u^\mu}{c^2} (u\cdot G \cdot s)\right) \frac{\widetilde bd}{ m}\;.
\end{align}
Two different constants $\widetilde a$ and $\widetilde b$ are introduced now since the two terms shown involving $F$ and $G$ tensors could be included in Schwinger solution independently with different constants. Intuition demands that $\widetilde a=\widetilde b$. However, aside from algebraic simplicity we do not find any compelling argument for this assumption. 

We return now to the definition of the $G$ tensor \req{SGforce} to obtain 
\begin{align}\label{Gs}
G^{\mu\nu}s_\nu
=&\left(
s_\nu s_\alpha\partial^\mu F^{\star\,{\nu\alpha}}-
s\cdot \partial F^{\star\,{\mu\alpha}}s_\alpha\right)\\ \notag
=&-s\cdot \partial F^{\star\,{\mu\nu}}s_\nu\;.
\end{align}
The first term in the first line vanishes by antisymmetry of $F^\star$ tensor. We also have
\begin{equation}\label{uGs}
u\cdot G\cdot s=-s\cdot \partial u\cdot F^\star\cdot s\;.
\end{equation}
Using \req{Gs} and \req{uGs} we can combine in \req{TBMT2} the first two terms in both lines, and the last terms in both lines to obtain 
\begin{align}\label{TBMT22}
\frac{ds^\mu}{d\tau}=&\frac{1+\widetilde a}{ m}\left(eF^{\mu\nu} -
\frac{1+\widetilde b}{1+\widetilde a}\;
s\cdot \partial\, F^{\star\,{\mu\nu}}d\,\right)s_\nu\\ \notag
-&\widetilde a\dsfrac{u^\mu}{mc^2} \left(u\cdot\left( eF -\dsfrac{\widetilde b}{\widetilde a}\; s\cdot \partial\, F^\star d\,\right)\cdot s\right)\;.
\end{align}
This equation agrees with \req{TBMTgradFinal} only when $\widetilde a=\widetilde b$. However, this requirement is neither mathematically nor physically necessary. For example using \req{SGforce200} we easily check $s\cdot \dot u+u\cdot \dot s=0$ without any assumptions about $\widetilde a, \widetilde b$. 

As \req{TBMT2} shows the physical difference between factors $\widetilde a$ and $\widetilde b$ is related to the nature of the interaction: the \lq magnetic\rq\ tensor $G$ is related to $\widetilde b$ only. Thus for a neutral particle $e\to 0$ we see in \req{TBMT22} that the torque depends only on $\widetilde b$. Conversely, when the effect of magnetic potential is negligible \req{TBMT22} becomes the textbook spin dynamics that depends on $\widetilde a$ alone.

To make further contact with textbook physics we note that the coefficient of the first term in \req{TBMT22} 
\begin{equation}\label{mustMU}
\frac{1+\widetilde a}{ m}e=2(1+\widetilde a)\frac{e\hbar}{2 m}\,\frac 1 \hbar=\widetilde g\, \mu_\mathrm{B}\,\frac 1 \hbar\;, \quad \widetilde g=2(1+\widetilde a)\;,
\end{equation}
should reproduce in leading order the torque coefficient in \req{tork} as is expected from study of quantum correspondence. However, quantum correspondence could mean $\widetilde a =a+a^2/2$, which follows comparing exact solutions of the Dirac equation with spin precession for the case we explored~\cite{AndrewSCoulomb} and which is not exactly the motion of a muon in storage ring. However, this means that in order to compare the measurement of magnetic moment of the muon carried out on macroscopic scale~\cite{Grange:2015fou,Bennett:2006fi} with quantum computations requires a further step, the establishment of quantum correspondence at the level of precision at which the anomaly is measured. 
 
\section{Search for variational principle action} \label{VarPrin} 
At the beginning of earlier discussions of a covariant extension to the Lorentz force describing the Stern-Gerlach force was always a well invented covariant action. However, the Lorentz force itself is not a consistent complement of the Maxwell equations. The existence of radiation means that an accelerated particle experiences radiation friction. The radiation-reaction force has not been incorporated into a variational principle~\cite{Rafelski:2017hyt,Hadad:2010mt}. Thus we should not expect that the Stern-Gerlach force must originate in a simple action. 

We seek a path $x^\mu(\tau)$ in space-time that a particle will take considering an action that is a functional of the 4-velocity $u^\mu(\tau)=dx^\mu/d\tau$ and spin $s^\mu(\tau)$. Variational principle requires an action $I(u,x;s)$. When $I$ respects space-time symmetries the magnitudes of particle mass and spin are preserved in the presence of electromagnetic (EM) fields. We also need to assure that $u^2=c^2$ which constrains the form of force and thus $I$ that is allowed. Moreover, we want to preserve gauge invariance of the resultant dynamics.

The component in the action that produced the LHS (inertia part) of the Lorentz force remains in discussion. To generate the Lorentz force one choice of action is
\begin{equation}\label{action0}
 I_\mathrm{Lz} (u,x)=-\!\int d\tau \;mc\sqrt{u^2} -e \!\int d\tau\, u(\tau)\cdot A(x(\tau)) \;.
\end{equation}
We note that reparametrization of $\tau \to k\tau$ considering $u=dx/d\tau$ has no effect on value of $I_\mathrm{Lz}$. 

Variation with respect to path lead to
\begin{equation}\label{variation}
\dsfrac{d}{d\tau}mc\frac{u^\mu}{\sqrt{u^2}}=L_\mathrm{Lz}^\mu= 
u_\nu\partial^\mu \!e A^\nu\! -\dsfrac{d\,eA^\mu}{d\tau}\;,
\end{equation}
where the RHS produces upon differentiation of $eA^\mu(x(\tau))$ the usual Lorentz force
\begin{equation}\label{LFusual}
L_\mathrm{Lz}^\mu= 
e (\partial^\mu A^\nu\! -\partial^\nu A^\mu)\, u_\nu=eF^{\mu\nu}u_\nu\;.
\end{equation}
Multiplying \req{variation} with $mc{u_\mu}/{\sqrt{u^2}}$ we establish by antisymmetry of the tensor $F^{\mu\nu}$ \req{LFusual} that also the product with the LHS in \req{variation} vanishes. This means that $(mc{u_\nu}/{\sqrt{u^2}})^2=m^2c^2\equiv p^2=Const.$ Henceforth
\begin{equation}\label{Momentum}
p^\mu\equiv mc\dsfrac{u^\mu}{\sqrt{u^2}}\;.
\end{equation}

There is a problem when we supplement in \req{action0} the usual action $I_\mathrm{Lz}$ by a term $I_\mathrm{m}$ based on our prior consideration of $A^\mu\to A^\mu+B^\mu$, see subsection~\ref{ssec:dipolepot}. The problem one encounters is that the quantity $B^\mu$ contains additional dependence on $s^\mu(\tau)$ which adds another term to the force. Let us look at the situation explicitly
\begin{equation}
\label{action}
 I (u,x\,;s)= I_\mathrm{Lz}+I_\mathrm{m}\;,\quad 
I_\mathrm{m}\equiv -\!\!\int\!\! d\tau\, u\cdot B(u,x\,;s)\, d \;.
\end{equation} 
Here the dependence on $s^\mu(\tau)$ is akin to a parameter dependence; some additional consideration defines the behavior, in our case this is the TBMT equations.

Varying with respect to the path the modified action \req{action} we find the modified covariant force 
\begin{equation}\label{LFfull}
\dsfrac{dp^\mu}{d\tau}=L_\mathrm{L}^\mu+ L_\mathrm{S1}^\mu+ L_\mathrm{S2}^\mu \;,
\end{equation}
with two new contributions 
\begin{align}\label{LFspin1}
L_\mathrm{S1}^\mu&= (\partial^\mu B^\nu-\partial^\nu B^\mu)u_\nu=G^{\mu\nu}u_\nu\;,\\ 
\label{LFspin2}
L_\mathrm{S2}^\mu&= - F^{\star\,{\mu\nu}}\,\dsfrac{ds_\nu}{d\tau}\, d\;. 
\end{align} 
We applied here with $A\to B$ the result seen in \req{variation}, and the additional term $L_\mathrm{S2}^\mu$ follows by remembering to take proper time derivative of $s^\mu$. The first term \req{LFspin1} is as we identified previously in \req{SGforce2}. We note that another additional term arises if and when an additional power of $\sqrt{u^2}$ to accompany $u\cdot B$ as was done in \cite{VanDam:1980aw}. An unsolved problem is created by the torque-like term, \req{LFspin2}. 

If we replace in our thoughts ${ds_\nu}/{d\tau}$ in \req{LFspin2} by the TBMT equation \req{TBMT} or as would be more appropriate by its extended version \req{TBMT2}, we see that the force $L_\mathrm{S2}^\mu$ would be quadratic in the fields containing also field derivatives. However, by assumption we modified the action limiting the new term in \req{action} to be linear in the fields and derivatives. Finding non linear terms we learn that this assumption was not justified. However, if we add the quadratic in fields term to the action we find following the chain of arguments just presented that a cubic term is also required and so on; with derivatives of fields appearing along. 

We have searched for some time for a form that avoids this circular conundrum, but akin to previous authors we did not find one. Clearly a \lq more\rq\ first principle approach would be needed to create a consistent variational principle based equation system. On the other hand we have presented before a formulation of spin dynamics which does not require a variational principle in the study the particle dynamics: as is we have obtained a dynamical equation system empirically. Our failing in the search for an underlying action is not critical. A precedent situation comes to mind here: the radiation emitted by accelerated charges introduces a \lq radiation friction\rq which must be studied~\cite{Rafelski:2017hyt,Hadad:2010mt} without an available action, based on empirical knowledge about the energy loss arising for accelerated charges.

\section{Experimental consequences}\label{Experiment}
\subsection{Non covariant form of dynamical equations}\label{NonCov}
\subsubsection{Laboratory frame}\label{ssec:Lab}
In most physical cases we create a particle guiding field which is at rest in the laboratory. Particle motion occurs with respect to this prescribed field and thus in nearly all situations it is practical to study particle position $z^\mu(\tau)$ in the laboratory frame of reference. Employing the Lorentz-coordinate transformations from the particle rest frame to the laboratory frame we obtain
\begin{align}
\label{noncovariantP}
\dsfrac{d\,z^\mu}{d\tau}\equiv u^\mu|_\mathrm{L}=&c \gamma\{1,\vec \beta\}\;,\qquad \vec \beta \equiv\dsfrac{d\vec z}{d\,ct}=\dsfrac{\vec v}{c}\;,\\[0.3cm]
\label{noncovariantS}
s^\mu|_\mathrm{L}=&\left\{\vec \gamma\vec\beta\cdot \vec s,\; \left(\dsfrac{\gamma}{\gamma+1}\,\gamma\,\vec \beta\cdot \vec s\right) \,\vec \beta+\vec s\right\}
\,,
\end{align}
where as usual $\gamma=1/\sqrt{1-\beta^2}$ and one often sees the spin written with $\gamma^2/(\gamma+1)=(\gamma-1)/\beta^2$. 

One easily checks that \req{noncovariantP} and \req{noncovariantS} also satisfy \req{constraint}: $u_\mu s^\mu=0$. A classic result of TBMT reported in textbooks is that the longitudinal polarization $\hat {\vec \beta}\cdot \vec s$ for $g\simeq 2$ and $\beta\to 1$ is a constant of motion. This shows that for a relativistic particle the magnitude of both time-like and space-like components of the spin 4-vector \req{noncovariantS} can be arbitrarily large, even if the magnitude of the 4-vector is bounded $s_\mu s^\mu=-\vec s^{\,2}$. This behavior parallels the behavior of 4-velocity $u^\mu u_\mu=c^2 $. 

We remind that to obtain in the laboratory frame the usual Lorentz force we use the 4-velocity with respect to the Laboratory frame \req{noncovariantP}, with laboratory defined tensor $F$, \ie\ with laboratory given $\EE,\;\BB$ EM-fields
\begin{equation}\label{triviaF}
 \dsfrac{d (m u^\mu |_\mathrm{L})}{d\tau}=\left(eF^{\mu\nu}u_\nu\right)|_\mathrm{L}
=eF^{\mu\nu}|_\mathrm{L}\;u_\nu|_\mathrm{L}\;.
\end{equation}
Sometimes it is of advantage to transform \req{triviaF} to the particle rest frame. Such a transformation $L$ with $Lu|_\mathrm{rest}=u_\mathrm{L}$ when used on the left hand side in \req{triviaF} produces proper time differentiation of the transformation operator, see also~\cite{Lobanov:1999}. Such transformation into a co-rotating frame of reference originates the Thomas precession term in particle rest frame for the torque equation. This term is naturally present in covariant formulation when we work in the laboratory reference frame. 

For the full force \req{SGforce200} we thus have
\begin{align}\label{SGforce201}
 \dsfrac{d (m u^\mu |_\mathrm{L})}{d\tau}
=& eF^{\mu\nu}|_\mathrm{L}u_\nu|_\mathrm{L}\\
 -& d \;s_\alpha|_\mathrm{L}\;
\left( \partial^\alpha F^{\star\,{\mu\nu}}\right)|_\mathrm{L}\; u_\nu|_\mathrm{L}\;.
\end{align}
We see that in laboratory frame of reference a covariant gradient of the fields is prescribed, \ie\ that some apparatus prescribes the magnitude 
\begin{equation}\label{gradEM}
Q^{\alpha\mu\nu}|_\mathrm{L}\equiv \partial^\alpha F^{*\,\mu\nu}|_\mathrm{L}\;,
\end{equation}
which allows for a moving particle with $u^\mu|_\mathrm{L}$ \req{noncovariantP} and $s^\mu|_\mathrm{L}$ \req{noncovariantS} to experience the Stern-Gerlach force $F^\mu_\mathrm{SG}$
\begin{equation}\label{SGLab}
F^\mu_\mathrm{SG}|_\mathrm{L}\equiv - d\,s_\alpha|_\mathrm{L} Q^{\alpha\mu\nu}|_\mathrm{L}u_\nu|_\mathrm{L}\;.
\end{equation}
We have gone to extraordinary length in arguing \req{SGLab} to make sure that the forthcoming finding of the Lorentz boost of field inhomogeneity is not questioned.

\subsubsection{Magnetic potential in the laboratory frame}\label{ssec:BinLab}
We evaluate in the laboratory frame the form of \req{genB2}. The computation is particularly simple once we first recall the laboratory format of the Lorentz force $F^\mu_\mathrm{L}$
 \begin{equation}\label{LorentzLab}
F^\mu_\mathrm{L}|_\mathrm{L}= F^{\mu\nu}(x)u_\nu|_\mathrm{L}=
c\gamma\left\{\vec \beta \cdot \EE/c, \;\EE/c+\vec \beta\times \BB\right\} 
\end{equation} 
The magnetic part of the action will be evaluated (see second line below) in analogy to above. We now consider 
 \begin{align}\label{uBLab}
B\cdot u|_\mathrm{L}=&u\cdot F^\star\cdot s|_\mathrm{L}
=-s^\mu|_\mathrm{L}\;(F^\star_{\mu\nu}u^\nu)|_\mathrm{L}\\ \notag
=&-s^\nu|_\mathrm{L}\;c \gamma \left\{-\vec \beta \cdot \BB, \; \BB-\vec \beta\times \EE/c\right\} \\ \notag
=&c\gamma\left(
\vec\beta \cdot \vec s\;\vec \beta\cdot \BB\frac{\gamma}{\gamma+1}
-\vec s\cdot( \BB-\vec \beta\times \EE/c)
\right)
\end{align}
where we used in 2nd line i) $F^\star_{\mu\nu}$ follows from the usual $F^{\mu\nu}$ upon exchange of $\EE/c\leftrightarrow\BB$ and ii) flip $\vec \beta \to -\vec \beta $ to account for contravariant and not covariant 4-velocity. In the 3rd line we used $\gamma( \gamma /(\gamma+1) -1)=-\gamma/(\gamma+1)$. Notable in \req{uBLab} is the absence of the highest power $\gamma^2$ as all terms cancel, the result is linear in (large) $ \gamma$. 

For the magnetic action potential energy of a particle in lab frame we obtain
 \begin{align}\label{duBLab}
U\equiv & \; B\cdot u|_\mathrm{L}d
= \;\gamma\left(
K\hat{\vec\beta} \cdot \vec \mu \;\hat{\vec \beta}\cdot \BB 
-\vec \mu\cdot (\BB-\vec \beta\times \EE/c)
\right)\;,\\[-0.3cm] 
\notag
K=&\;\beta^2\;\dsfrac{\gamma}{\gamma+1}=1-\sqrt{1-\beta^2}=\left\{\begin{array}{lr}
\frac 1 2 \beta^2\;, & \mathrm{for}\ \beta\to 0\\[0.2cm]
1\;, & \mathrm{for}\ \beta\to 1
\end{array}
\right.
\end{align}
\Req{duBLab} extends the rest frame $\vec \beta=0$ \req{PotMu} and represents covariant generalization of \req{muB}. In ultrarelativistic limit all terms in \req{duBLab} have the same magnitude. 

\subsubsection{Field to particle energy transfer}\label{ssec:EfieldTOP}

We now consider the energy gain by a particle per unit of laboratory time, that is we study the 0th component of \req{SGforce200}
\begin{align}\label{Egain}
\dsfrac{dE}{dt}=&c \dsfrac{d\tau}{dt} \dsfrac{d (m u^0 |_\mathrm{L})}{d\tau}
=c\gamma^{-1}\widetilde F^{0\nu}u_\nu|_\mathrm{L}\\ \notag
=&e\EE\cdot \vec v+
cd \;s^\alpha|_\mathrm{L}(\partial_\alpha\BB)|_\mathrm{L}\cdot \vec v \\ \label{EgainExp}
\dsfrac{dE}{dt}=&(e\EE+(\vec\mu\cdot\vec \nabla)\BB)\cdot \vec v\\ \notag
&\hspace*{0.5cm}+ \gamma\vec \beta\cdot\vec \mu \left(\dsfrac{\partial \BB}{c\partial t}
 +\dsfrac{\gamma}{\gamma+1}(\vec\beta\cdot\vec \nabla)\;\BB\right)\cdot\vec v\;.
\end{align}
A further simplification is achieved considering
\begin{equation}
\dsfrac{\partial \BB}{c\partial t} +(\vec\beta\cdot\vec \nabla)\;\BB=
\dsfrac{\partial \BB}{c\partial t} +\sum_{i=1}^3 \dsfrac{dx_i}{cdt}\dsfrac{\partial\BB}{\partial x_i}
=\dsfrac{d\BB}{c\,dt}\;, 
\end{equation}
where the total derivative with respect to time accounts for both, the change in time of the laboratory given field $\BB$, and the change due to change of position in the field by the moving particle. We thus find two parts
\begin{align}\label{Egain2}
\dsfrac{dE}{dt}=& \vec v\cdot\left(e\EE+(\vec\mu\cdot\vec \nabla)\BB
-K\hat{\vec \beta}\cdot\vec \mu \; (\hat{\vec\beta}\cdot\vec \nabla)\;\BB\right)
\\ \notag
 +&\vec \beta \cdot \dsfrac{d\BB}{ dt} \;\gamma \; \vec \beta\cdot\vec \mu\;,
\end{align}
where the 2nd line is of particular interest as it is proportional to $\gamma$. Focusing our attention on this last term: we can use $\vec\beta=c\vec p/E$ and $\gamma\vec\beta=\vec p/mc$. Upon multiplication with $E$ and remembering that $c^2\vec p d\vec p=E dE$ we obtain 
\begin{equation}
\vec p\cdot\left(\dsfrac{d\vec p}{dt}- \dsfrac{d\BB}{ dt}\dsfrac{\vec \mu\cdot \vec p}{mc^2}\right)=0\;.
\end{equation}
which in qualitative terms implies an exponential response of particle momentum as it crosses a magnetic field
\begin{equation}
|\vec p|\simeq mc\; e^{\pm (|\BB|-\BB_0))|\vec \mu|/mc^2}\;.
\end{equation}
However, even a magnetar magnetic field of up $10^{11}\mathrm{T}$ will not suffice to impact electron momentum decisively in view of the smallness of the electron magnetic moment $ 5.8 10^{-11}$\;$\mathrm{MeV/T}$. However, in ultrarelativistic heavy ion collisions at LHC a 10,000 stronger very non-homogeneous \BB-fields arise.

\subsection{Neutral particle hit by a light pulse}\label{PlaneWM}
\subsubsection{Properties of equations}
The dynamical equations developed here have a considerably more complex form compared to the Lorentz force and TBMT spin precession in constant fields~\cite{Lobanov:1999}. We need field gradients in the Stern-Gerlach force, and in the related correction in the TBMT equations. Since the new physics appears only in the presence of a particle magnetic moment, we simplify by considering neutral particles. We now show that the external field described by a light wave (pulse) lends itself to an analytical solution effort. This context could be of practical relevance in the study of laser interaction with magnetic atoms, molecules, the neutron and maybe neutrinos.

For $e=0$ our equations \req{SGforce200} and \req{TBMT22} read
\begin{align}\label{eq:firstdynam}
\dot{u}^\mu =& - s\cdot\partial F^{*\mu\nu}u_\nu\,\frac{d}{m}\;,\\
\label{eq:seconddynam}
\dot{s}^\mu =& -s\cdot\partial F^{*\mu\nu}s_\nu\frac{1+\widetilde{b}}{m}\,d + u^\mu u \cdot (s\cdot \partial) F^* \cdot s \,\frac{\widetilde{b}\, d}{mc^2}\;.
\end{align}
The external light wave field is a pulse with
\begin{equation}\label{LPulse}
A^\mu = \varepsilon^\mu f(\xi)\;, \quad \quad \xi = k \cdot x\;, \quad k\cdot \varepsilon=0\;.
\end{equation}
The derivative of the dual EM tensor for linear fixed in space pulse polarization $\varepsilon^\mu$ is
\begin{align}\label{eq:emtensor}
(s\cdot \partial)F^{*\mu\nu}
 =& (k\cdot s) \epsilon^{\mu\nu\alpha\beta}k_\alpha\varepsilon_\beta f^{\prime\prime}(\xi)\;,
\end{align}
prime \lq $\prime$\rq\ indicates derivative with respect to the phase $\xi$. 

Notice that if we contract \req{eq:emtensor} with $k_\mu$ or $\varepsilon_\mu$ we get zero because Levi-Civita tensor $\epsilon^{\mu\nu\alpha\beta}$ is totally antisymmetric. Therefore contracting \req{eq:firstdynam} with either $k_\mu$ or $\varepsilon_\mu$ we find
\begin{align}\label{eq:conserveKU}
0=&k\cdot \dot u\ \to\ k\cdot u =k\cdot u(0)\;,\quad u^\mu(0)=u^\mu(\tau_0)\\ 
\label{eq:conservePU}
0=& \varepsilon \cdot \dot u\ \to\ \varepsilon \cdot u =\varepsilon \cdot u(0)\;.
\end{align}
 We further note that the argument of the light pulse \req{LPulse} satisfies 
\begin{equation}\label{LPulse2}
\xi = k \cdot x \ \to\ \dot \xi= k \cdot \dot x=k \cdot u=k \cdot u(0)\;.
\end{equation}
where we used \req{eq:conserveKU}. Thus we conclude that the particle follows the pulse such that 
\begin{equation}\label{LPulse2b}
\xi = k \cdot x=\tau\; k \cdot u(0)+\xi_0\;, \quad \xi_0= k \cdot x(0)\;.
\end{equation}
The two conservation laws \req{eq:conserveKU} and \req{eq:conservePU} along with \req{LPulse2} make the light pulse an interesting example amenable to an analytical solution. 

We now evaluate several invariants in the laboratory frame seeking understanding of their relevance. A particle moving in the laboratory frame in consideration of \req{noncovariantP} experiences in its rest frame a plane wave with the Doppler shifted frequency
\begin{equation}\label{DopplerS}
k \cdot u(0)=\gamma_0(1-\vec n\vec\cdot\vec \beta_0)\omega\; 
\end{equation}
which is unbounded as it grows with particle laboratory Lorentz-$\gamma_0$. However, $k\cdot s$, the projection of spin onto plane wave 4-momentum $k^\mu$ is bounded. To see this we recall the constraint \req{constraint} which in the laboratory frame reads 
\begin{equation}\label{uslab}
S^0_\mathrm{L}-\vec \beta\cdot \vec S_\mathrm{L}=0\;.
\end{equation}
We thus obtain
\begin{equation}\label{kslab}
k\cdot s(\tau)=k\cdot s(\tau)|_\mathrm{L}=
|\vec k|\left(S^0_\mathrm{L}-\vec n\cdot \vec S_\mathrm{L}\right)
=|\vec k|(\vec \beta-\vec n)\cdot \vec S_\mathrm{L}\;,
\end{equation}
where we used \req{uslab} in last equality. Since $\vec \beta$ and $\vec n=\vec k/|\vec k|$ are unit-magnitude vectors we find 
\begin{equation}\label{kslab2}
(k\cdot s(\tau))^2\le 4 \vec k^{\,2}\; \vec S_\mathrm{L}^{\,2}\;.
\end{equation}
The magnitude of the spin vector in the lab frame is constrained by \req{RFrame1} 
\begin{equation}\label{sslab}
-\vec s^{\,2}=S^{0\,2}_\mathrm{L}-\vec S_\mathrm{L}^{\,2}=
(\vec \beta\cdot \vec S_\mathrm{L})^2-\vec S_\mathrm{L}^{\,2}=-\sin^2\!\theta\; \vec S_\mathrm{L}^{\,2}\;,
\end{equation}
where we again used \req{uslab}. Combining \req{kslab2} and \req{sslab} we see that except when particle is moving exactly in direction of $S_\mathrm{L}$ ($\sin^2\!\theta=0$), the magnitude of $(k\cdot s(\tau))^2$ is bounded. 

\subsubsection{Invariant acceleration and spin precession} 
Even without knowing the explicit form for $u^\mu(\tau),\;s^\mu(\tau)$ we were able to obtain~\cite{FormanekSOl} the invariant acceleration
\begin{equation}\label{invacc}
\dot{u}^2(\tau) = - \left(\frac{d}{m} f^{\prime\prime}(\xi(\tau))\; k \cdot s(\tau)\; k \cdot u(0)\right)^{\!2}\;.
\end{equation}
This result follows using the usual trick of taking a further (proper) time derivative of \req{eq:firstdynam} (multiplied by a suitable factor) and on RHS eliminating $\dot u$ by using \req{eq:firstdynam}. Multiplying the result with $u_\mu$ and eliminating $u \cdot \ddot{u}$ using the 2nd differential of $u^2=c^2$ produces \req{invacc}.

We see in \req{invacc} that the magnitude of the 4-force created by a light pulse and acting on an ultrarelativistic particle is dependent on square of the product of the 2nd derivative of pulse function with respect to $\xi$, $f^{\prime\prime}(\xi)$, with the Doppler shifted frequency \req{DopplerS}. The value \req{invacc} is negative since acceleration is a space-like vector. 

As we discussed below \req{sslab} the spin precession factor $k\cdot s$ seen in \req{invacc} is bounded. We were able to obtain a soluble formulation of the spin precession dynamics described by the dimensionless variable
\begin{equation}\label{invacc3}
y=k\cdot s(\tau)\dsfrac{\widetilde bd}{mc\,C_1}
\end{equation}
which satisfies the differential equation 
\begin{equation}\label{invacc32}
\left(\dsfrac{d\,y(s)}{d\,s}\right)^2 =y^2(1-y^2)\; \quad 
s=\left(f^\prime(\xi(\tau))-f^\prime(\xi_0)\right) C_1
\end{equation}
obtained performing suitable manipulations of dynamical equations prior to solving for $u^\mu(\tau),\;s^\mu(\tau)$. We are seeking bounded periodic solutions of nonlinear \req{invacc32} no matter how large the constant $C_1$ determined by the initial conditions
\begin{align}\label{ksK1}
C_1 \equiv &\frac{\widetilde{b}\,d}{mc}\;k\cdot s(0)\;C_2\;,\qquad C_2\ge 1\;,\\[0.2cm] 
\label{ksK2}
C_2 \equiv & \left.\sqrt{\frac{|(k\cdot u)^2|s^2|-[(k\cdot u)(\varepsilon \cdot s) - (\varepsilon \cdot u)(k \cdot s) ]^2|}{c^2(k \cdot s)^2}}\right|_{\tau=0}\;.
\end{align}
$C_2$ contains the initial particle Lorentz-$\gamma$ factor. One can see several possible solutions of interest of \req{invacc32}; for example $y=\sin(\phi(s))$ satisfies all constraints. It leads to the pendulum type differential equation and we recognize that high intensity light pulses can flip particle spin. However, there are other relevant solutions, \eg\ $y\propto 1/\cosh z$.

Upon solution of \req{invacc32} $k\cdot s(\tau)$ is known, given \req{LPulse2b} we also know the dependence of \req{eq:emtensor} on proper time $\tau$. Hence \req{eq:firstdynam} can be solved for $u^\mu$ and \req{eq:seconddynam} can be solved for $s^\mu$ resulting in an analytical solution of the dynamics of a neutral magnetic dipole moment in the field of a light pulse of arbitrary shape. The full description of the dynamics exceeds in length this presentation and will follow~\cite{FormanekSOl}.
 
\section{Conclusions}\label{conclude}
 
The Stern-Gerlach covariant extension of the Lorentz force has seen considerable interest as there are many immediate applications listed in first paragraph. Here we have:\\
1) introduced in \req{RFrame} the covariant classical 4-spin vector $s^\mu$ in a way expected in the context of Poincare symmetry of space-time;\\
2) presented a unique linear in fields form of the covariant magnetic moment potential, \req{genB2}, which leads to a natural generalization of the Lorentz force;\\
3) shown that the resultant Amperian, \req{Aforce1}, and Gilbertian, \req{SGforce200}, forms of the magnetic moment force are equivalent;\\
4) extended the TBMT torque dynamics, \req{TBMT2}, making these consistent with the modifications of the Lorentz force;\\
5) demonstrated the need to connect the magnetic moment magnitude entering the Stern-Gerlach force with the one seen in the context of torque dynamics, subsection~\ref{ssec:TBMTgradNU};\\
6) shown that variational principle based dynamics has systemic failings when both position and spin are addressed within present day conceptual framework, see section~\ref{VarPrin};\\
7) reduced the covariant dynamical equations to laboratory frame of reference uncovering important features governing the coupled dynamics, see section~\ref{NonCov};\\
8) obtained work done by variations of magnetic field in space-time on a particle, \req{Egain2};\\
9) shown salient features of solutions of neutral particles with non-zero magnetic moment hit by a laser pulse, see section~\ref{PlaneWM};\\
 

\end{document}